\documentclass[final,3p,times,twocolumn]{elsarticle}

\pdfoutput=1 

 \usepackage{graphics}
 \usepackage{graphicx}
\usepackage{subfigure}
  \usepackage{lineno}
\usepackage{amssymb}

\usepackage{mathtools}
\usepackage{amsmath}
\usepackage{xspace}
\usepackage{macross}
\usepackage[hyperindex,breaklinks]{hyperref} 

\usepackage{preprintcover}
\PreprintCoverPaperTitle{Measurement of the $WW$ cross section in $\sqrt{s} = 7$~TeV \\ $pp$
collisions with the ATLAS detector and \\ limits on anomalous gauge couplings}  

\PreprintIdNumber{CERN-PH-EP-2012-060}  

\PreprintCoverAbstract{This Letter reports a measurement of the $WW$ production cross section 
in $\sqrt{s} = 7$~TeV $pp$ collisions using data corresponding to an integrated luminosity of 
1.02~fb$^{-1}$ collected with the ATLAS detector.  Using leptonic decays of oppositely charged 
$W$ bosons, the total measured cross section is $\sigma(pp\rightarrow WW) = 
54.4 \pm 4.0~{\rm (stat.)}~\pm 3.9~{\rm (syst.)}~\pm 2.0~{\rm (lumi.)}$~pb, consistent with the 
Standard Model prediction of $\sigma(pp\rightarrow WW) = 44.4 \pm 2.8$~pb.  Limits on anomalous 
electroweak triple-gauge couplings are extracted from a fit to the transverse-momentum distribution 
of the leading charged lepton in the event. }

\PreprintJournalName{Physics Letters B}  

\journal{Physics Letters B}

\begin{document}

\begin{frontmatter}

\title{Measurement of the $WW$ cross section in $\sqrt{s} = 7$~TeV $pp$
collisions with the ATLAS detector and limits on anomalous gauge couplings}

\author{The ATLAS Collaboration}

\address{}

\begin{abstract}
This Letter reports a measurement of the $WW$ production cross section in $\sqrt{s} = 7$~TeV $pp$ 
collisions using data corresponding to an integrated luminosity of 1.02~fb$^{-1}$ collected with 
the ATLAS detector.  Using leptonic decays of oppositely charged $W$ bosons, the total measured cross 
section is $\sigma(pp\rightarrow WW) = 54.4 \pm 4.0~{\rm (stat.)}~\pm 3.9~{\rm (syst.)}~\pm 
2.0~{\rm (lumi.)}$~pb, consistent with the Standard Model prediction of $\sigma(pp\rightarrow WW) 
= 44.4 \pm 2.8$~pb.  Limits on anomalous electroweak triple-gauge couplings are extracted from a 
fit to the transverse-momentum distribution of the leading charged lepton in the event.  
\end{abstract}


\end{frontmatter}

\section{Introduction}
\label{intro}
Measurements of $WW$ production at the LHC provide important tests of the Standard Model (SM), 
in particular of the $WWZ$ and $WW\gamma$ triple gauge couplings (TGCs) resulting from the 
non-Abelian nature of the {\it SU(2)}$_L \times U(1)_Y$ symmetry group.  Precise measurements 
of TGCs are sensitive probes of new physics in the electroweak sector and are complementary to 
direct searches.  Furthermore, since $WW$ production is a background to possible new processes 
such as the production of the SM Higgs boson, a precise measurement of the $WW$ cross section 
is an important step in the search for new physics.

This Letter describes the measurements of the $WW$ cross section and of TGCs in $pp$ collisions 
at $\sqrt{s} = 7$ TeV.  The dominant SM $WW$ production mechanisms are $s$-channel and $t$-channel 
quark-antiquark annihilation, with a 3\% contribution from gluon-gluon fusion.  The cross section 
is measured in the fiducial phase space of the detector using $WW \rightarrow l\nu l\nu$ decays in 
final states with electrons and muons, and is extrapolated to the total phase space.  The fiducial 
phase space includes geometric and kinematic acceptance.  The total production cross section of 
oppositely charged $W$ bosons is measured according to the equation \cite{wzxsec}

\begin{equation}
\label{eq:xsec}
\sigma(pp\rightarrow WW) = \frac{N_{\mathrm{data}} - N_{\mathrm{bg}}}{A_{WW}C_{WW}{\cal{L}}
{\cal{B}}},
\end{equation}
\noindent
where $N_{\mathrm{data}}$ and $N_{\mathrm{bg}}$ are the number of observed data events and 
estimated background events, respectively, $A_{WW}$ is the kinematic and geometric acceptance, 
$C_{WW}$ is the ratio of the number of measured events to the number of events produced in the 
fiducial phase space, ${\cal{L}}$ is the integrated luminosity of the data sample, and ${\cal{B}}$ 
is the branching ratio for both $W$ bosons to decay to $e\nu$ or $\mu\nu$ (including decays 
through tau leptons with additional neutrinos).  The fiducial cross section is defined as 
$\sigma \times A_{WW} \times {\cal{B}}$ \cite{wzxsec}.

Previous measurements of $WW$ production using the CMS and ATLAS detectors, both based on
the data recorded in 2010 and corresponding to an integrated luminosity of 36 pb$^{-1}$, have 
found $\sigma(pp \rightarrow WW) = 41.1 \pm 15.3~{\rm (stat.)} \pm 5.8~{\rm (syst.)} \pm 
4.5~{\rm (lumi.)}$ pb \cite{cmsww} and $\sigma(pp \rightarrow WW) = 41^{+20}_{-16}~{\rm (stat.)} 
\pm 5~{\rm (syst.)} \pm 1~{\rm (lumi.)}$ pb \cite{atlasww}, respectively.  CMS has additionally 
used these data to set limits on anomalous gauge-coupling parameters at higher center of mass 
energies than corresponding measurements at the Tevatron \cite{tev} and LEP \cite{lep}.

\section{ATLAS detector}
 \label{detector}
The ATLAS detector \cite{atlas} consists of an inner tracking system (inner detector, or ID) 
surrounded by a superconducting solenoid providing a 2 T magnetic field, electromagnetic 
and hadronic calorimeters, and a muon spectrometer (MS) incorporating three large 
superconducting toroid magnets arranged with an eight-fold azimuthal coil symmetry around the 
calorimeters.  The ID consists of silicon pixel and microstrip detectors, surrounded by a 
transition radiation tracker.  The electromagnetic calorimeter is a lead/liquid-argon (LAr) 
detector.  Hadron calorimetry is based on two different detector technologies, with scintillator 
tiles or LAr as active media, and with either steel, copper, or tungsten as the absorber material.  
The MS comprises three layers of chambers for the trigger and for track measurements.  

A three-level trigger system is used to select events.  The level-1 trigger is implemented 
in hardware and uses a subset of detector information to reduce the event rate to a design 
value of at most 75 kHz.  This is followed by two software-based trigger levels, level-2 and 
the event filter, which together reduce the event rate to about 200 Hz recorded for analysis.  

The nominal $pp$ interaction point at the centre of the detector is defined as the origin of a 
right-handed coordinate system.  The positive $x$-axis is defined by the direction from the 
interaction point to the centre of the LHC ring, with the positive $y$-axis pointing upwards, 
while the $z$-axis is along the beam direction.  The azimuthal angle $\phi$ is measured around 
the beam axis and the polar angle $\theta$ is the angle from the $z$-axis.  The pseudorapidity 
is defined as $\eta = -\ln \tan(\theta/2)$.

\section{Data sample and event selection}
\label{data}
The data used for this analysis correspond to an integrated luminosity of $1.02 \pm 0.04$~fb$^{-1}$ 
\cite{lumi}, recorded between April and June of 2011.  Events are selected with triggers requiring 
either a single electron with $\pt > 20$ GeV and $|\eta| < 2.5$ or a single muon with $\pt > 18$ GeV 
and $|\eta| < 2.4$.  Additional data collected with a trigger requiring a single muon with $\pt > 40$ 
GeV, $|\eta| < 1.05$, and looser identification criteria are used to increase efficiency.  The 
combination of triggers results in $\approx 100\%~(98\%)$ trigger efficiency for events with $WW$ 
decays to $e\nu \mu\nu$ and $e\nu e\nu$ ($\mu\nu \mu\nu$) passing the selection described below. 

The $WW$ event selection begins with the identification of electrons and muons, requiring 
exactly two of these particles with opposite charge.  Electrons are reconstructed with a 
clustering algorithm in the electromagnetic calorimeter and matched to an ID track.  To 
distinguish electrons from hadrons, selection criteria \cite{eid} are applied based on the 
quality of the position and momentum match between the extrapolated track and the calorimeter 
cluster, the consistency of the longitudinal and lateral shower profiles with an incident 
electron, and the observed transition radiation in the TRT.  Electrons are required to lie 
within the fiducial regions of the calorimeters ($|\eta| < 1.37$ or $1.52 < |\eta| < 2.47$), 
have $\pt > 25$ GeV ($\pt > 20$ GeV for the lower \pt electron in the $e\nu e\nu$ decay 
channel), and be isolated in the calorimeter and tracker.  Calorimeter isolation requires the 
summed transverse energies deposited in calorimeter cells, excluding those belonging to the 
electron cluster, in a cone of radius $\Delta R = \sqrt{(\Delta\eta)^2 + (\Delta\phi)^2} = 0.3$ 
around the electron direction to be $< 4$ GeV.  Tracker isolation requires the summed \pt of ID 
tracks in a cone of radius $\Delta R = 0.2$ centered on and excluding the electron track to be 
$<10\%$ of the electron \pt. 

The muon reconstruction algorithm begins with a track from the MS to determine the muon's $\eta$, 
and then combines it with an ID track to determine the muon's momentum \cite{muonreco}.  Muons are 
required to have $\pt > 20$ GeV and $|\eta| < 2.4$, and in the $\mu\nu \mu\nu$ channel at least 
one muon must have $\pt > 25$ GeV.  Decays of hadrons to muons are suppressed using calorimeter 
and track isolation.  The calorimeter isolation requires the summed transverse energies deposited 
in calorimeter cells in a cone of radius $\Delta R = 0.2$ around the muon track to be less than 
$15\%$ of the muon's $p_{\mathrm T}$.  The track isolation requirement is the same as for electrons.  
The tracks associated with muon and electron candidates must have longitudinal and transverse impact 
parameters consistent with originating from the primary reconstructed vertex.  The primary vertex 
is defined as the vertex with the highest $\sum \pt^2$ of associated ID tracks. 

The presence of neutrinos is characterized by an imbalance of transverse momentum in the event.  
The missing transverse momentum (\met) is the modulus of the event $-\vec{p}_{\mathrm T}$ vector, 
calculated by summing the transverse momentum determined from each calorimeter cell's energy and 
direction with respect to the primary vertex.  Cells with $|\eta| < 4.5$ are used in the 
calculation and a correction is applied to account for the momentum of measured muons.

Misreconstructed leptons and jets, as well as leptons from tau decays, are suppressed by applying 
cuts on $\met\times\sin\Delta\phi$ when $\Delta\phi < \pi/2$.  Here, $\Delta\phi$ is the azimuthal 
angle between the missing transverse momentum and the nearest charged lepton or jet; small 
$\Delta\phi$ indicates that \met is dominated by a mismeasured lepton or jet, or by the presence 
of neutrinos in the direction of the lepton or jet, as would occur in a tau decay.  The lower 
cuts on \met, or $\met\times\sin\Delta\phi$ for $\Delta\phi < \pi/2$, are 25 GeV in the $e\nu\mu\nu$ 
channel, 40 GeV in the $e\nu e\nu$ channel, and 45 GeV in the $\mu\nu\mu\nu$ channel.  The thresholds 
in the $e\nu e\nu$ and $\mu\nu \mu\nu$ channels are more stringent than in the $e\nu\mu\nu$ channel 
to suppress the background from Drell-Yan (DY) production of $ee$ and $\mu\mu$ pairs.

Background from top-quark production is rejected by vetoing events containing a reconstructed 
jet with $\pt > 25$ GeV and $|\eta| < 4.5$.  Jets are reconstructed with the anti-$k_{\mathrm t}$ 
algorithm \cite{antikt} with a radius parameter of $R = 0.4$.  A further 30\% reduction of top-quark 
background is achieved by rejecting events with a jet with $\pt > 20$ GeV, $|\eta| < 2.5$, and 
identified as originating from a $b$-quark ($b$-jet).  The identification of $b$-jets combines 
information from the impact parameters and the reconstructed vertices of tracks within the jet 
\cite{bquarks}.  The additional $b$-jet rejection reduces $WW$ acceptance by 1.3\%.
 
Resonances with dilepton decays are removed by requiring $ee$ and $\mu\mu$ invariant masses to 
be greater than 15 GeV and not within 15 GeV of the $Z$-boson mass.  To suppress backgrounds from 
heavy-flavour hadron decays, events with an $e\mu$ invariant mass below 10 GeV are also removed.  
The complete event selection yields 202 $e\nu\mu\nu$, 59 $e\nu e\nu$, and 64 $\mu\nu\mu\nu$ 
candidates. 

\section{Background estimation}
\label{background}
The selected data sample contains $26 \pm 3$\% background to the $WW$ production process 
(Table~\ref{tbl:backgrounds}).  In decreasing order of size, the main background processes 
are: DY production of dileptons, with significant \met arising from misreconstructed jet(s) or 
charged lepton(s); $t\bar{t}$ and $tWb$ production, where the $b$-quarks in the $WWb\bar{b}$ 
final state are not rejected by the jet veto; $(W \rightarrow l\nu)$ + jet, where the jet 
is misidentified as a lepton; $WZ \rightarrow l\nu ll$ production, where one lepton is not 
reconstructed; $(W \rightarrow l\nu) + \gamma$, where the photon converts in the inner detector 
and is misreconstructed as an electron; $ZZ \rightarrow ll\nu\nu$ production; and cosmic-ray 
muons overlapping a $pp$ collision (which is negligible).  

Backgrounds are estimated using a combination of Monte Carlo (MC) samples including a full 
{\sc geant} \cite{geant} simulation of the ATLAS detector \cite{atlassim}, and control samples 
(independent of the measurement sample) from data.  The simulation includes the modeling of 
multiple $pp$ interactions in the same bunch crossing (pile-up), as well as corrections 
(determined from data) to improve the modeling of reconstructed objects.

\begin{table*}[t]
\caption{The estimated background event yields in the selected $WW$ data sample.  The 
first uncertainty is statistical, the second systematic. }
\begin{center}
\begin{tabular}{cccc}
\hline 
Production process & $e\nu\mu\nu$ selection & $e\nu e\nu$ selection & $\mu\nu\mu\nu$ selection 
\\
\hline 
DY               & $13.0 \pm 2.1 \pm 1.6$ & $12.5 \pm 2.3 \pm 1.4$ & $10.9 \pm 2.5 \pm 1.4$
\\
Top              & $11.9 \pm 1.8 \pm 2.4$ & $ 3.1 \pm 0.5 \pm 0.6$ & $ 3.8 \pm 0.6 \pm 0.8$
\\
$W$ + jet        & $10.0 \pm 1.6 \pm 2.1$ & $ 4.1 \pm 1.3 \pm 0.9$ & $ 4.2 \pm 1.1 \pm 1.3$
\\
Diboson          & $ 5.1 \pm 1.0 \pm 0.7$ & $ 2.1 \pm 0.8 \pm 0.3$ & $ 2.9 \pm 0.4 \pm 0.4$
\\
\hline
Total background & $40.0 \pm 3.3 \pm 3.6$ & $21.7 \pm 2.8 \pm 1.8$ & $21.8 \pm 2.8 \pm 2.1$ 
\\ 
\hline
\end{tabular}
\end{center}
\label{tbl:backgrounds}
\end{table*}

The DY background is estimated using the {\sc alpgen} \cite{alpgen} Monte Carlo generator 
interfaced to {\sc pythia} \cite{pythia} for parton showering.  To test the modeling of 
\met, data are compared to simulated $Z/\gamma^*$ events where the lepton pair forms an 
invariant mass within 15 GeV of the $Z$-boson mass.  The DY MC accurately models the number 
of events above the thresholds on \met or $\met\times\sin\Delta\phi$ used to select $WW$ 
events, after subtracting the $\approx 20\%$ non-DY contributions.  A $12\%$ relative 
systematic uncertainty on the DY prediction is taken from the statistical precision of the 
MC validation in the control sample.

\begin{figure}[!ht]
 \begin{center}
 \includegraphics[width=0.485\textwidth]{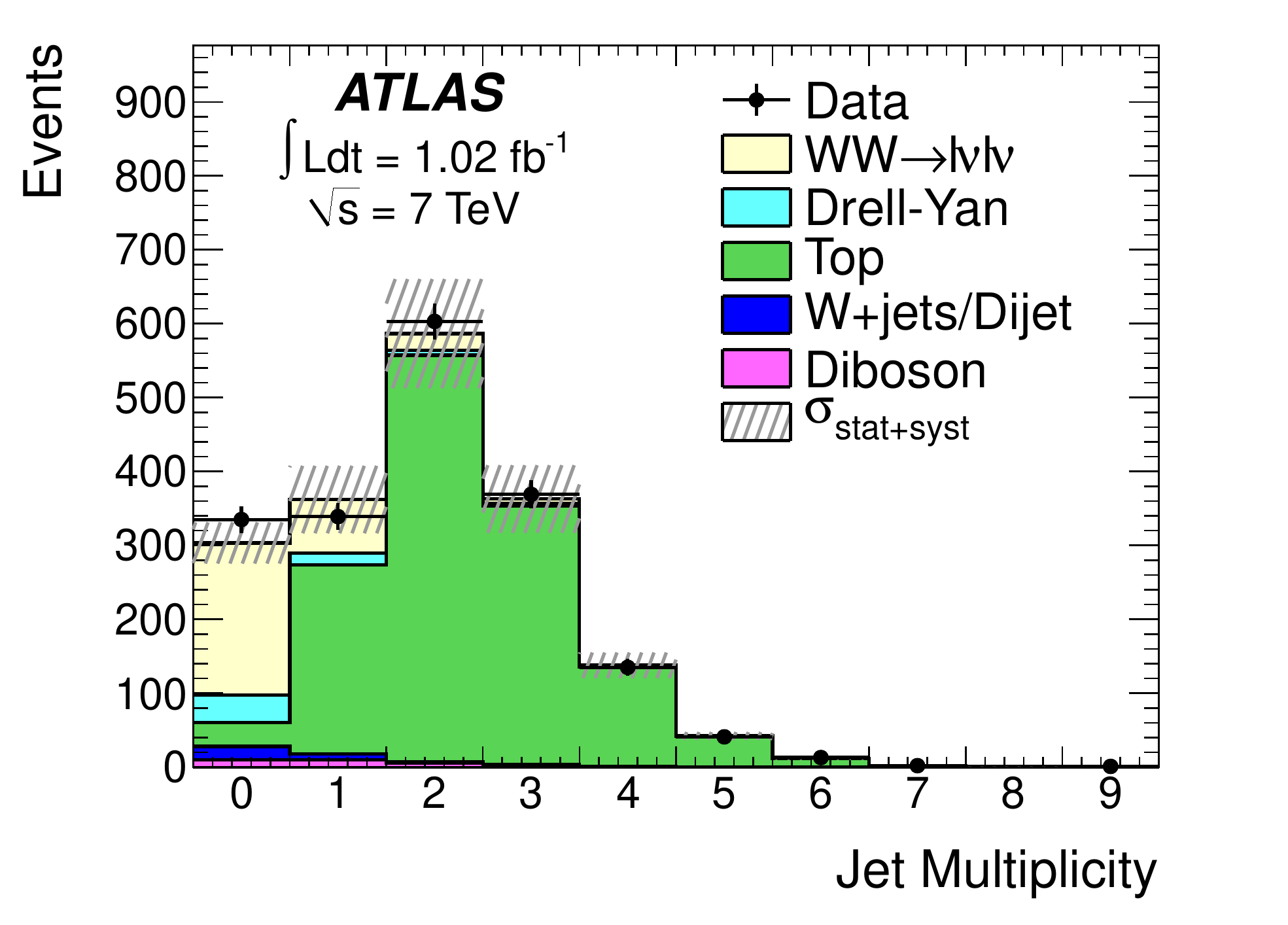}
 \caption{The multiplicity distribution of jets with $p_T > 25$ GeV for the combined dilepton 
channels, after all $WW$ selection cuts except the jet veto requirement.  The systematic 
uncertainties shown in the $>$0-jet bins include only those on the integrated luminosity and 
the theoretical cross sections. }
 \label{fig:njet}
 \end{center}
 \end{figure}

Background from top-quark production arises when the final-state $b$-quarks have low transverse 
momentum ($\pt < 20$ GeV), are not identified as $b$-jets (for $20 < \pt < 25$ GeV), or are in 
the far forward region ($|\eta| > 4.5$).  To model this background, {\sc mc@nlo} \cite{mcatnlo} 
samples of $t\bar{t}$ and {\sc AcerMC} \cite{acermc} samples of $tWb$ production are used, 
respectively, with corrections derived from the data.  An overall normalization factor is 
determined from the ratio of events in data to those predicted by the top-quark MC using the 
$WW$ selection without any jet rejection.  This sample is dominated by top-quark decays, as 
shown in Fig.~\ref{fig:njet}; a 24\% contribution from other processes is subtracted in the 
normalization.  The subtraction of the $WW$ component is based on the SM prediction of $WW$ 
production, with an uncertainty that covers the difference between the prediction and the 
cross section measurement reported in this Letter.  The relative cross sections of $t\bar{t}$ 
to $tWb$ are set by the generator calculations of $\sigma = 164.6$ pb and $\sigma = 15.6$ pb, 
respectively.
 
A key aspect of the top-quark background prediction is the modeling of the jet veto acceptance.  
To reduce the associated uncertainties, a data-based correction is derived 
using a top-quark-dominated sample based on the $WW$ selection but with the requirement of at 
least one $b$-jet with $p_T > 25$ GeV \cite{topcorrections}.  In this sample, the 
ratio $P_1$ of events with one jet to the total number of events is sensitive to the modeling 
of the jet energy spectrum in top-quark events.  A multiplicative correction based on the ratio 
$P_1^{data}/P_1^{MC}$ is applied to reduce the uncertainties resulting from the jet veto 
requirement.  The residual uncertainty on the background prediction due 
to jet energy scale and resolution is small (4\%) compared to uncertainties from the $b$-quark 
identification efficiency (6\%), parton shower modeling (12\%), statistical uncertainty on the 
$P_1^{\mathrm{data}}/P_1^{MC}$-based correction (12\%), and unmodeled $t\bar{t}$-$tWb$ 
interference and higher order QCD corrections (15\%).  As a cross-check, the normalization 
of the top-quark background is extracted from a fit to the jet multiplicity distribution; the 
result is consistent with the primary estimate.

The $W$ + jet process contributes to the selected sample when one or more hadrons in the jet 
decay to, or are misidentified as, a charged lepton.  Jets reconstructed as electrons or muons 
predominantly arise from misidentification or heavy-flavour quark decays, respectively.  This 
background is estimated with a pass-to-fail ratio $f_e$ ($f_{\mu}$), defined as the ratio of 
the number of electron (muon) candidates passing the electron (muon) identification criteria to 
the number of candidates failing the criteria.  These ratios are measured in data samples 
dominated by hadronic jets collected with a trigger requiring an electromagnetic cluster or a 
muon candidate.  All candidates are required to pass a loose set of selection criteria, 
including an isolation requirement.  The measured $f_e$ and $f_{\mu}$ are then applied as 
multiplicative factors to events satisfying all $WW$ selection cuts except with one lepton 
failing the identification criteria but passing the looser criteria.

The above procedure measures $f_e$ and $f_{\mu}$ ratios averaged over misidentified jets and
heavy-flavour quark decays in jet-dominated samples.  If, for example, the ratio $f_e$ differs 
for these two contributions, the $W$ + jet prediction could be biased.  To address this issue, 
two sets of loose criteria are applied to electron candidates, one based on the track and the 
shower profile and expected to enhance the misidentification fraction, and the other based on 
the isolation and expected to enhance the heavy flavour fraction.  The $f_e$ ratio is measured 
for these criteria separately in events where there is an additional $b$-jet and events where 
there is no such jet.  From the combination of measurements, the heavy-flavour and 
misidentification contributions are separated; the resulting $W$ + jet background is consistent 
with that obtained using the inclusive $f_e$ for the misidentification and heavy-flavour 
components.  A similar separation is not performed for $f_{\mu}$, since heavy-flavour decays 
dominate the contribution of background muons from the $W$ + jet process.

The systematic uncertainty on the $W$ + jet prediction is dominated by a 30\% variation of the 
ratios $f_e$ and $f_{\mu}$ with the jet $p_{\mathrm T}$ threshold.  This variation is sensitive to 
the relative fraction of quarks and gluons in the samples used to measure $f_e$ and $f_{\mu}$, and 
thus encompasses potential differences in $f_e$ and $f_{\mu}$ ratios between these samples and those 
used to estimate the $W$ + jet background.

Several alternative methods are used to check the $W$ + jet prediction and give consistent results.  
The first method applies the measured $f_e$ and $f_{\mu}$ ratios to an inclusive $W$ + jets data 
sample, and then determines the fraction of expected events with no additional jets using $W$ + jets 
Monte Carlo events with two identified leptons.  The second method defines different sets of ``loose'' 
lepton criteria and independently measures efficiencies for lepton identification and rates for 
misidentified or decaying hadrons to pass the standard identification criteria.  Background from 
dijet production is estimated with this method and is found to be small; it is implicitly included 
in the primary estimate.

Monte Carlo estimates of the $W\gamma$, $WZ$, and $ZZ$ backgrounds are obtained using a combination 
of {\sc alpgen} and {\sc pythia} (for $W\gamma$) and {\sc herwig} \cite{herwig} with {\sc jimmy} 
\cite{jimmy} (for the others), normalized to the next-to-leading order (NLO) cross sections 
calculated with {\sc mcfm} \cite{mcfm}.  The ${\cal{O}}$(10\%) systematic uncertainty on these 
backgrounds is dominated by the uncertainty on the jet energy scale.

\section{$WW$ acceptance modeling}
\label{wwaccept}
The $WW$ total cross section measurement requires the knowledge of the $A_{WW}$ and $C_{WW}$ 
factors given in Eq.~\ref{eq:xsec}.  The acceptance factor $A_{WW}$ is defined as the ratio of 
generated $WW$ events in the fiducial phase space to those in the total phase space.   The 
correction factor $C_{WW}$ is defined as the ratio of measured events to generator-level events 
in the fiducial phase space.  The value of this ratio is determined primarily by lepton trigger 
and identification efficiencies, with a small contribution from differences in generated and 
measured phase space due to detector resolutions.  The fiducial phase space is defined at 
generator level as:
\begin{itemize}
\item 
Muon $\pt > 20$ GeV and $|\eta| < 2.4$ ($\pt > 25$ GeV for at least one muon in 
the $\mu\nu\mu\nu$ channel); 
\item 
Electron $\pt > 20$ GeV and either $|\eta| < 1.37$ or $1.52 < |\eta| < 2.47$ ($\pt > 25$ GeV 
in the $e\nu\mu\nu$ channel and for at least one electron in the $e\nu e\nu$ channel);
\item 
No anti-$k_{\mathrm t}$ jet ($R = 0.4$) with $\pt > 25$ GeV, $|\eta| < 4.5$, and 
$\Delta R(e,$jet$) > 0.3$; 
\item 
No anti-$k_{\mathrm t}$ jet with $\pt > 20$ GeV, $|\eta| < 2.5$, $\Delta R(e,$jet$) > 0.3$, 
and $\Delta R(b,$jet$) < 0.3$, where the $b$-quark has $\pt > 5$ GeV;
\item 
Neutrino $|\sum \vec{p}_{\mathrm T}|$~or~$|\sum\vec{p}_{\mathrm T}| \times \sin\Delta\phi$ 
(for $\Delta\phi < \pi/2$) $> 45, 40, 25$~GeV in the $\mu\nu\mu\nu$, $e\nu e\nu$ and $e\nu \mu\nu$ 
channels, respectively ($\Delta\phi$ is the azimuthal angle between the neutrino 
$\sum \vec {p}_{\mathrm T}$ and the nearest charged lepton);
\item 
$m_{\ell\ell} > 15~(10)$ GeV in the $\mu\nu\mu\nu$ and $e\nu e\nu$ channels ($e\nu\mu\nu$ 
channel);
\item $|m_{\ell\ell} - m_Z| > 15$ GeV in the $\mu\nu\mu\nu$ and $e\nu e\nu$ channels, 
\end{itemize}

\noindent
where $m_Z$ is the $Z$ boson mass.  To reduce the dependence on the model of QED final-state 
radiation, the electron and muon \pt include contributions from photons within $\Delta R = 0.1$ 
of the lepton direction.

Estimates of $A_{WW}$ and $C_{WW}$ are based on samples of $q\bar{q}~\rightarrow~WW$ and 
$gg~\rightarrow~WW$ events generated with {\sc mc@nlo} and {\tt gg2WW} \cite{gg2ww}, respectively.  
Initial parton momenta are modeled with CTEQ 6.6 \cite{cteq} parton distribution functions (PDFs).  
The underlying event and parton showering are modeled with {\sc jimmy}, and hadronization and 
tau-lepton decays with {\sc herwig}.  Data-based corrections measured with $W$ and $Z$ boson data 
are applied to reduce uncertainties, as described below.  Because the corrections are applied to 
$WW$ MC samples, residual uncertainties on the fiducial cross section measurement are based on 
the kinematics of SM $WW$ production.

The combined factor $A_{WW}\times C_{WW}$ is estimated separately for each leptonic decay 
channel, including decays to tau leptons (Table~\ref{tbl:wwacc}).  Tau-lepton decays to 
hadrons are not included in the denominator for the acceptances in the table.  The impact 
of pile-up is modeled by adding {\sc pythia}-generated low-$Q^2$ events to the $WW$ MC 
according to the distribution of the number of additional collisions in the same bunch 
crossing in the data.  Effects on detector response from nearby bunches are also modeled 
using this distribution.

\begin{table*}[ht]
\caption{The total $WW$ acceptance $A_{WW}\times C_{WW}$ in the individual decay channels, and 
the expected number of SM $WW$ events ($N_{WW}$) for an integrated luminosity of 1.02 fb$^{-1}$. }
\begin{center}
\begin{tabular}{ccccccc}
\hline 
           & \multicolumn{2}{c}{$e\nu\mu\nu$ selection} 
           & \multicolumn{2}{c}{$e\nu e\nu$ selection}  
           & \multicolumn{2}{c}{$\mu\nu\mu\nu$ selection} 
\\
           & $WW \rightarrow e\nu\mu\nu$  & $WW \rightarrow l\nu \tau\nu$ 
           & $WW \rightarrow e\nu e\nu$   & $WW \rightarrow e\nu \tau\nu$ 
           & $WW \rightarrow \mu\nu\mu\nu$& $WW \rightarrow \mu\nu\tau\nu$  
\\
\hline 
$A_{WW}\times C_{WW}$ & $10.8\%$ & $3.0\%$ & $4.4\%$ & $1.1\%$ & $7.6\%$ & $1.6\%$ \\ 
$N_{WW}$              & $114.9$  & $12.0$  & $23.4$  & $2.3$   & $40.3$  & $3.3$   \\
\hline
\end{tabular}
\end{center}
\label{tbl:wwacc}
\end{table*}

A correction to the $q\bar{q} \rightarrow WW$ MC modeling of the jet veto is derived using 
$Z$-boson data.  The fraction of $Z$-boson events with no additional jets is compared between 
data and {\sc mc@nlo} simulated samples.  The ratio of this fraction in data to the fraction 
in the MC is applied as a multiplicative correction factor of 0.963 to the $WW$ MC.  The 
correction reduces the uncertainties due to jet energy scale and resolution to 1.1\%.  A 
theoretical uncertainty of 5.0\% on the jet veto acceptance contributes the largest uncertainty 
to $A_{WW}$, as shown in Table~\ref{tbl:accuncertainties}.

Contributions to \met include energy from the interacting protons' remnants (the underlying event), 
and from pile-up.  The dominant uncertainty arises from the detector response to the underlying 
event, and is evaluated by varying the individual calorimeter cell energy deposits in the MC 
\cite{dmet}.  To determine the uncertainty due to additional $pp$ interactions in the same bunch 
crossing as the hard-scattering process, the event $\vec{p}_{\rm T}$ measured with the calorimeter 
is compared between data and MC in $Z\rightarrow \mu\mu$ events.  The mean $|\vec{p}_{\rm T}|$ as 
a function of the number of reconstructed vertices agrees to within 3\% between data and MC, 
yielding a negligible uncertainty on the $WW$ acceptance.  The effect of collisions from other 
bunch crossings is studied by splitting $Z$-boson samples in data and MC according to the bunch 
position in the LHC train, and by smearing \met in the simulation samples to match the acceptance 
of a given \met cut in the data samples.  The resulting uncertainty on the $WW$ acceptance is small.

The efficiencies for triggering, reconstructing, and identifying charged leptons are measured 
as a function of lepton \pt and $\eta$ using $Z$ boson events and (for electrons) $W$ boson 
events \cite{wzxsec}.  Corrections to the MC derived from these data are within 1\% of unity for 
trigger and muon identification efficiencies and deviate from unity by up to 11\% at low $p_T$ for 
the electron identification efficiency.  Uncertainties on the corrections are largely due to the 
limited number of events available for the measurements and, in case of electron identification, 
from the estimate of the jet background contamination.

\begin{table*}[!ht]
\caption{Relative uncertainties, in percent, on the estimate of the product $A_{WW}\times C_{WW}$ 
for the individual $WW$ decay channels.  The uncertainty on $A_{WW}$ ($C_{WW}$) receives 
contributions from the last three (first six) sources. }
\begin{center}
\begin{tabular}{cccc}
\hline 
 & \multicolumn{3}{c}{Relative uncertainty (\%)} \\
Source of uncertainty
           & $e\nu\mu\nu$ selection 
           & $e\nu e\nu$ selection  
           & $\mu\nu\mu\nu$ selection 
\\
\hline 
Trigger efficiency               & 1.0 & 1.0 & 1.0 \\
Lepton efficiency                & 2.3 & 4.1 & 1.8 \\ 
Lepton \pt scale and resolution  & 0.4 & 1.0 & 0.1 \\ 
\met modeling                    & 0.6 & 1.0 & 2.2 \\ 
Jet energy scale and resolution  & 1.1 & 1.1 & 1.1 \\ 
Lepton acceptance                & 2.0 & 2.1 & 1.6 \\
Jet veto acceptance              & 5.0 & 5.0 & 5.0 \\ 
PDFs                             & 1.4 & 1.2 & 1.2 \\
\hline
Total                            & 6.2 & 7.2 & 6.2 \\
\hline
\end{tabular}
\end{center}
\label{tbl:accuncertainties}
\end{table*}

Finally, there are small uncertainties on the $WW$ production model.  Uncertainties on PDFs are 
determined using the CTEQ eigenvectors and the acceptance differences between the CTEQ 6.6 
and MSTW 2008 PDF sets \cite{mstw}.  The impact of unmodeled higher-order contributions is estimated 
by varying the renormalization and factorization scales coherently by factors of 2 and 1/2.  

The total acceptance uncertainty on the three channels combined is 6.2\%.

\section{Cross section results}
\label{xsec}
The $WW$ cross section is measured in the fiducial phase space and extrapolated to the total 
phase space.  The total cross section is defined in Eq.~\ref{eq:xsec}, while the fiducial cross 
section is 

\begin{equation}
\sigma_{\rm fid} = \frac{N_{\mathrm{data}} - N_{\mathrm{bg}}}{{\cal{L}} C_{WW}}. 
\end{equation}

\noindent
Uncertainties on the fiducial cross section measurement result from modeling lepton and 
jet efficiency, energy scale and resolution, and \met (the first five rows of 
Table~\ref{tbl:accuncertainties}).  Small uncertainties of 1.4\% ($\mu\nu\mu\nu$ and 
$e\nu e\nu$ channels) and 0.5\% ($e\nu\mu\nu$ channel) arise from the impact of QCD 
renormalization and factorization scale variations on lepton momenta (included in the 
sixth row of Table~\ref{tbl:accuncertainties}).  Table~\ref{tbl:results} shows $C_{WW}$ 
and the other components of the cross section measurements for each channel.  The 
measurements are performed by minimizing a likelihood fit to the observed data with respect 
to the $WW$ and background predictions for the three channels combined.  The measured cross 
sections are consistent with the SM predictions, differing by $+1.7\sigma$ ($e\nu\mu\nu$ 
channel), $+1.3\sigma$ ($e\nu e\nu$ channel) and $-0.1\sigma$ ($\mu\nu\mu\nu$ channel).  
Contributions from a hypothetical SM Higgs boson would be small:  2.9, 0.9, and 1.8 events 
in the $e\nu\mu\nu$, $e\nu e\nu$ and $\mu\nu\mu\nu$ channels, respectively, for a Higgs 
boson mass of 125 GeV.
 
\begin{table*}[!ht]
\caption{The measured total ($\sigma(pp\rightarrow WW)$) and fiducial ($\sigma_{\rm fid}$) cross 
sections and the components used in the calculations, as well as the SM predictions for the fiducial 
cross sections ($\sigma^{\rm SM}_{\rm fid}$).  The first uncertainty is statistical and the second 
systematic.  The 3.7\% relative uncertainty on the integrated luminosity is the third uncertainty on 
the measured cross sections.  The uncertainties on $\sigma^{\rm SM}_{\rm fid}$ are highly correlated 
between the channels. } 
\begin{center}
\begin{tabular}{cccc}
\hline 
                 & $e\nu\mu\nu$ selection   
                 & $e\nu e\nu$ selection    
                 & $\mu\nu\mu\nu$ selection 
\\
\hline 
Data             & 202  & 59  & 64 
\\
Background       & $40.0 \pm 3.3 \pm 3.6$ 
                 & $21.7 \pm 2.8 \pm 1.8$ 
                 & $21.8 \pm 2.8 \pm 2.1$ 
\\ 
$C_{WW}$         & $0.541 \pm 0.005 \pm 0.022$ 
                 & $0.396 \pm 0.005 \pm 0.019$ 
                 & $0.721 \pm 0.005 \pm 0.025$ 
\\
$A_{WW}$         & $0.161 \pm 0.001 \pm 0.008$ 
                 & $0.089 \pm 0.001 \pm 0.005$ 
                 & $0.082 \pm 0.001 \pm 0.004$ 
\\
$A_{WW} \times C_{WW}$ & $0.087 \pm 0.001 \pm 0.005$ 
                       & $0.035 \pm 0.001 \pm 0.003$ 
                       & $0.059 \pm 0.001 \pm 0.004$ 
\\
\hline
$\sigma(pp\rightarrow WW)$ [pb] & $56.3 \pm 4.9 \pm 3.9 \pm 2.1$ 
                    & $64.1 \pm 13.0 \pm 7.4 \pm 2.4$ 
                    & $43.2 \pm 8.1 \pm 4.5 \pm 1.6$ 
\\
$\sigma_{\rm fid}$ [fb] & $294 \pm 26 \pm 15 \pm 11$ 
                        & $92.0 \pm 18.9 \pm 9.4 \pm 3.4$ 
                        & $57.2 \pm 10.8 \pm 5.2 \pm 2.1$ 
\\
$\sigma^{\rm SM}_{\rm fid}$ [fb] & $230 \pm 19$
                                 & $63.4 \pm 5.3$
                                 & $59.0 \pm 4.7$
\\
\hline
\end{tabular}
\end{center}
\label{tbl:results}
\end{table*}

\begin{figure*}[!ht]
 \begin{center}
 \includegraphics[width=0.485\textwidth]{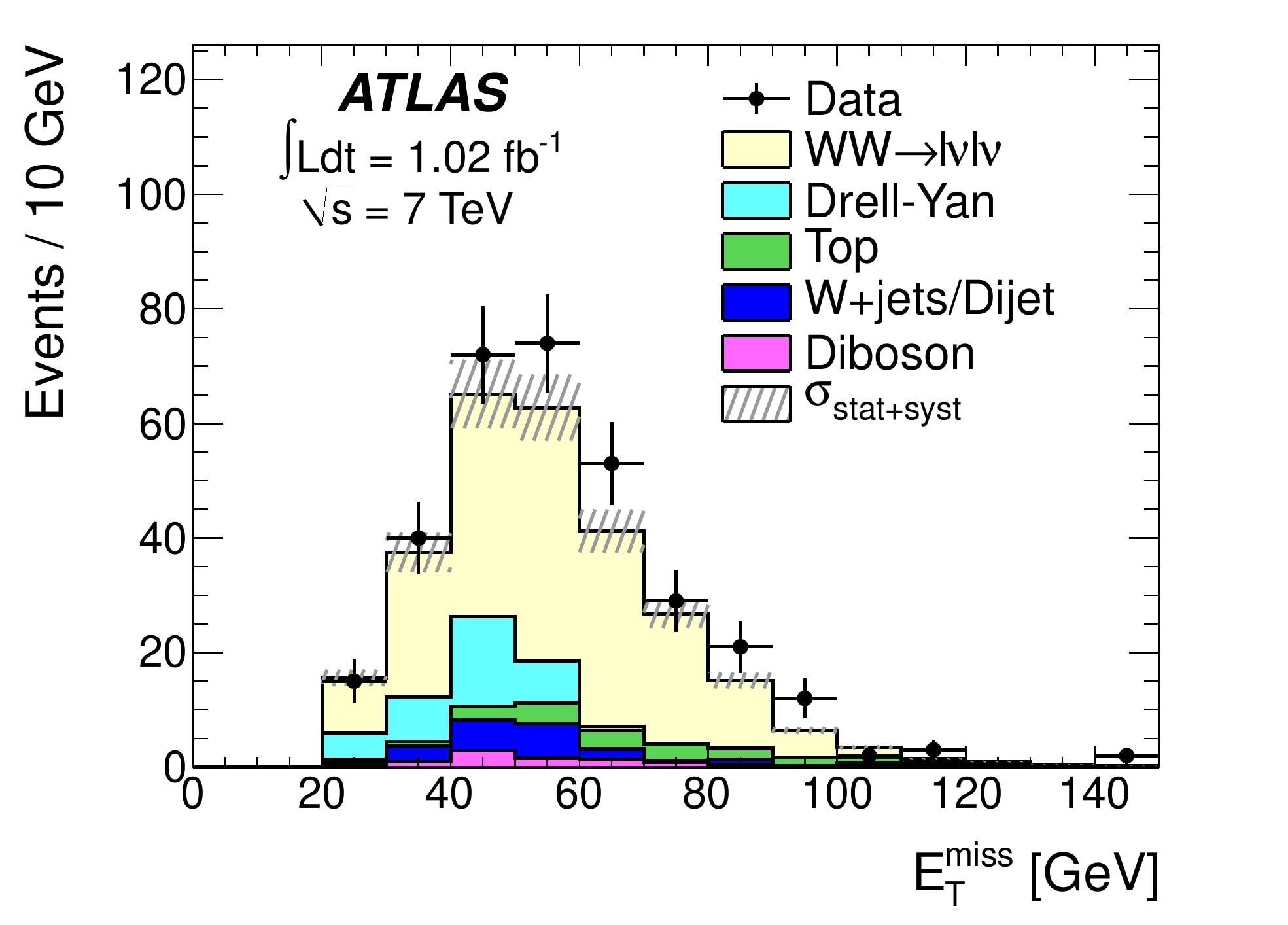}
 \includegraphics[width=0.485\textwidth]{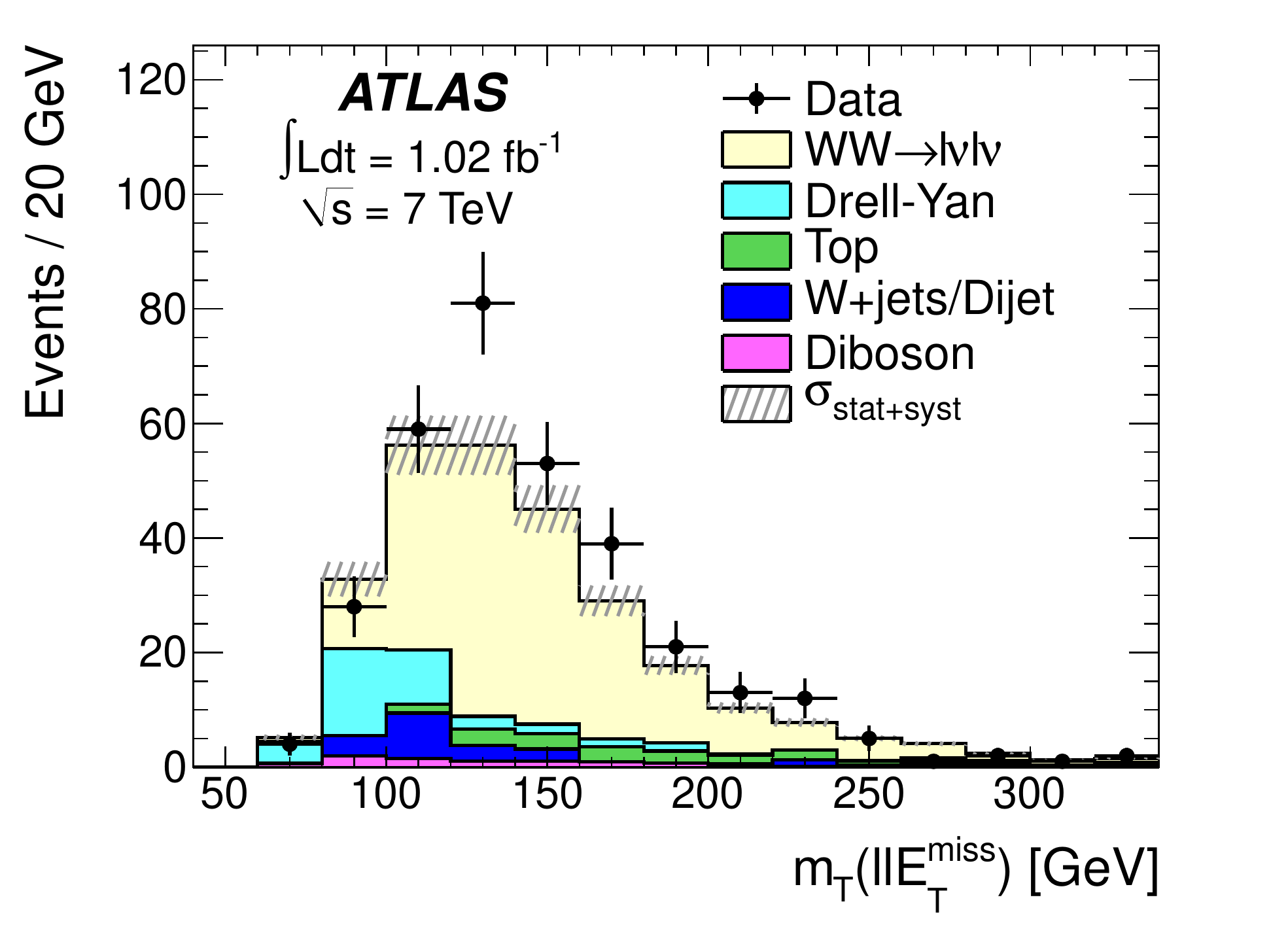}
 \includegraphics[width=0.485\textwidth]{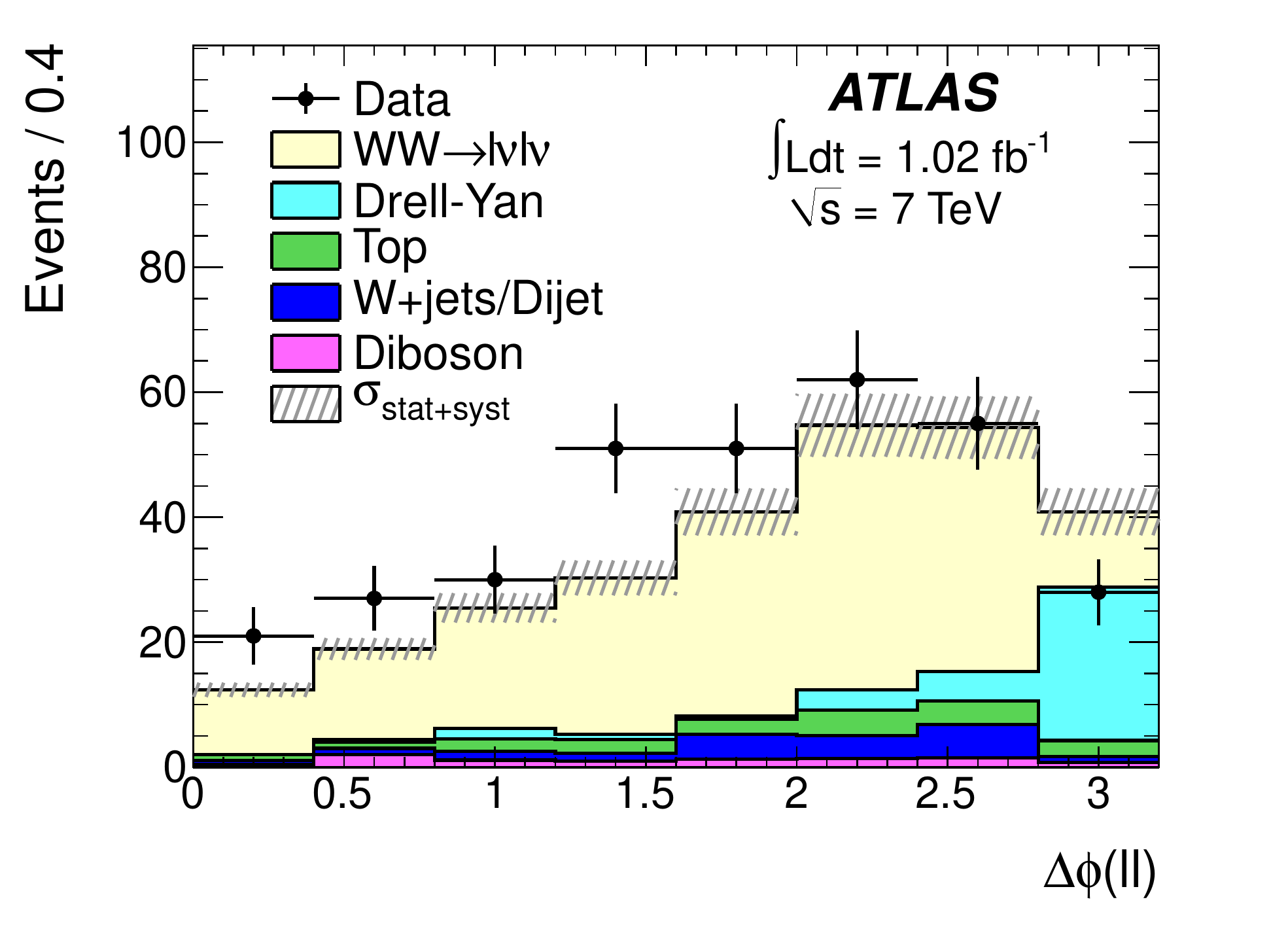}
 \includegraphics[width=0.485\textwidth]{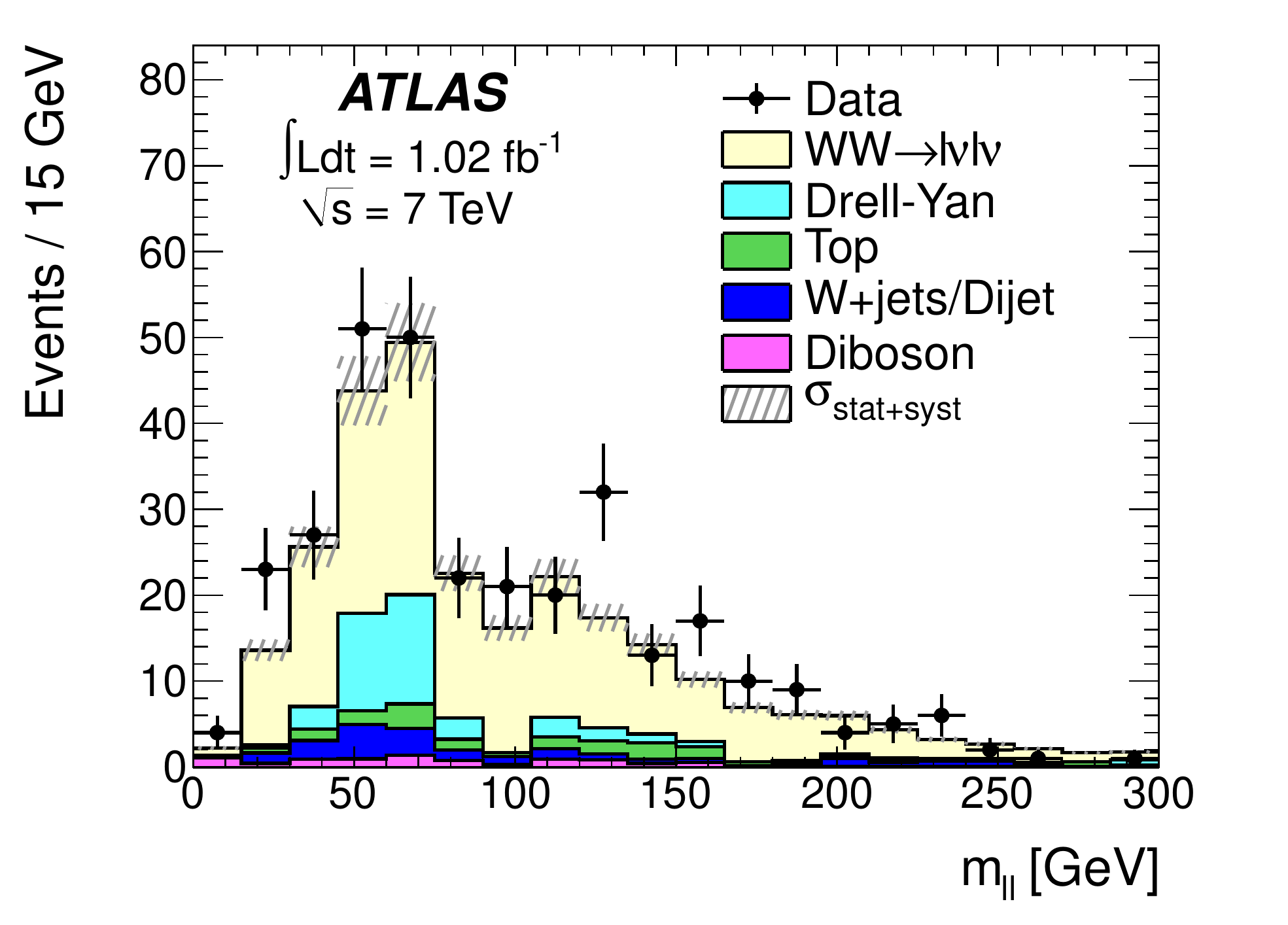}
 \caption{The \met (top left), $m_{\mathrm T}$ (top right), $\Delta\phi(l,l)$ (bottom left) 
and $m_{\ell\ell}$ (bottom right) distributions for the combined dilepton channels after all 
selection requirements.  The data (dots) are compared to the expectation from $WW$ and the 
backgrounds (histograms).  The $W$ + jet and dijet backgrounds are estimated using data.  
The hashed region shows the $\pm 1\sigma$ uncertainty band on the expectation.  }
 \label{fig:distributions}
 \end{center}
 \end{figure*}

The $A_{WW}$ uncertainty comes from PDFs and scale variations affecting the lepton and 
jet veto acceptances (the last three rows of Table~\ref{tbl:accuncertainties}).  The 
combined $A_{WW} \times C_{WW}$ and the total measured cross section in each channel 
are shown in Table~\ref{tbl:results}.  The contribution of leptons from tau decays
is included.  The channels are combined by maximizing a log likelihood, yielding 

\begin{align}
\sigma(pp\rightarrow WW)  = &\notag \\
  54.4 \pm 4.0~{\rm (stat.)} & \pm 3.9~{\rm (syst.)} \pm 2.0~{\rm (lumi.)}~{\rm pb}, \nonumber 
\end{align}

\noindent
to be compared with the NLO SM prediction of $\sigma(pp\rightarrow WW) = 44.4 \pm 2.8$ pb 
\cite{mcatnlo, gg2ww}.  Figure~\ref{fig:distributions} shows the following distributions 
for data and MC: \met, transverse mass, the azimuthal angle between the charged leptons 
[$\Delta\phi(l,l)$], and invariant mass of the charged leptons $[m_{\ell\ell}]$.  The 
transverse mass is $m_{\mathrm T}(ll\met) = \sqrt{(\pt^{l_1} + \pt^{l_2} + \met)^2 - 
\sum (p_i^{l_1} + p_i^{l_2} + E_i^{\mathrm{miss}})^2}$, where the sum runs over the $x$ and 
$y$ coordinates and $l_1$ and $l_2$ refer to the two charged leptons.

\section{Anomalous triple-gauge couplings}
\label{aTGC}
The $s$-channel production of $WW$ events occurs via the triple-gauge couplings $WW\gamma$ 
and $WWZ$.  Contributions to these couplings from new physics processes at a high energy 
scale would affect the measured cross section, particularly at high momentum transfer 
\cite{lagrangian}.  Below the energy scale of these new physics processes, an effective 
Lagrangian can be used to describe the effect of non-SM processes on the $WWV$ ($V=\gamma, Z$) 
couplings.  Assuming the dominant non-SM contributions conserve $C$ and $P$, the general 
Lagrangian for $WWV$ couplings is 

\begin{eqnarray}
{\cal{L}}_{WWV}/g_{WWV}&= ig_1^V(W^{\dagger}_{\mu\nu}W^{\mu}V^{\nu} - 
  W^{\dagger}_{\mu}V_{\nu}W^{\mu\nu}) + \nonumber  \\
 & i\kappa_V W^{\dagger}_{\mu} W_{\nu} V^{\mu\nu} + 
      \frac{i\lambda_V}{m_W^2} W^{\dagger}_{\lambda\mu}W^{\mu}_{\nu}V^{\nu\lambda}, 
\end{eqnarray}

\noindent
where $g_{WW\gamma} = -e,~g_{WWZ} = -e\cot\theta_W,~V_{\mu\nu} = \partial_{\mu} 
V_{\nu} - \partial_{\nu} V_{\mu}$ and $W_{\mu\nu} = \partial_{\mu} W_{\nu} - 
\partial_{\nu} W_{\mu}$.  The SM couplings are $g_1^V = \kappa_V = 1$ and 
$\lambda_V = 0$.  Individually, non-zero couplings lead to divergent cross 
sections at high $\sqrt{s}$, and non-SM values of the $g_1^V$ or $\kappa_V$ 
couplings break the gauge cancellation of processes at high momentum transfer.  
To regulate this behavior, a suppression factor depending on a scale $\Lambda$ 
with the general form

\begin{equation}
\lambda(\hat{s}) = \frac{\lambda}{(1 + \hat{s} / \Lambda^2)^2},
\end{equation}

\noindent
is defined for $\lambda$, $\Delta g_1 \equiv g_1 - 1$ and $\Delta \kappa \equiv \kappa - 1$.  
Here, $\lambda$ is the coupling value at low energy and $\sqrt{\hat{s}}$ is the 
invariant mass of the $WW$ pair.  The $g_1^{\gamma}$ coupling is fixed to its SM 
value by electromagnetic gauge invariance.

To reduce the number of $WWV$ coupling parameters, three specific scenarios are 
considered.  The first is the ``LEP scenario'' \cite{atgc, lepatgc}, where anomalous 
couplings arise from dimension-6 operators and electroweak symmetry breaking occurs 
via a light SM Higgs boson.  This leads to the relations

\begin{equation}
\Delta\kappa_{\gamma} = -\frac{\cos^2\theta_W}{\sin^2\theta_W} (\Delta \kappa_Z - 
\Delta g_1^Z)~{\rm and}~\lambda_{\gamma} = \lambda_Z,
\end{equation}

\noindent
leaving three free parameters ($\Delta g_1^Z,~\Delta\kappa_Z,~\lambda_Z$).  The 
parameter space can be further reduced by requiring equal couplings of the SU(2) 
and U(1) gauge bosons to the Higgs boson in the dimension-6 operators.  This adds 
the constraint $\Delta g_1^Z = \Delta \kappa_{\gamma}/(2\cos^2\theta_W)$ and is 
referred to as the ``HISZ scenario'' \cite{atgc}.  A third ``Equal Coupling scenario'' 
assumes common couplings for the $WWZ$ and $WW\gamma$ vertices ($\Delta\kappa_Z = 
\Delta\kappa_{\gamma},~\lambda_Z = \lambda_{\gamma}$, $\Delta g_1^Z = 
\Delta g_1^{\gamma} = 0$). 

\begin{figure}[!t]
\begin{center}
\includegraphics[width=0.485\textwidth]{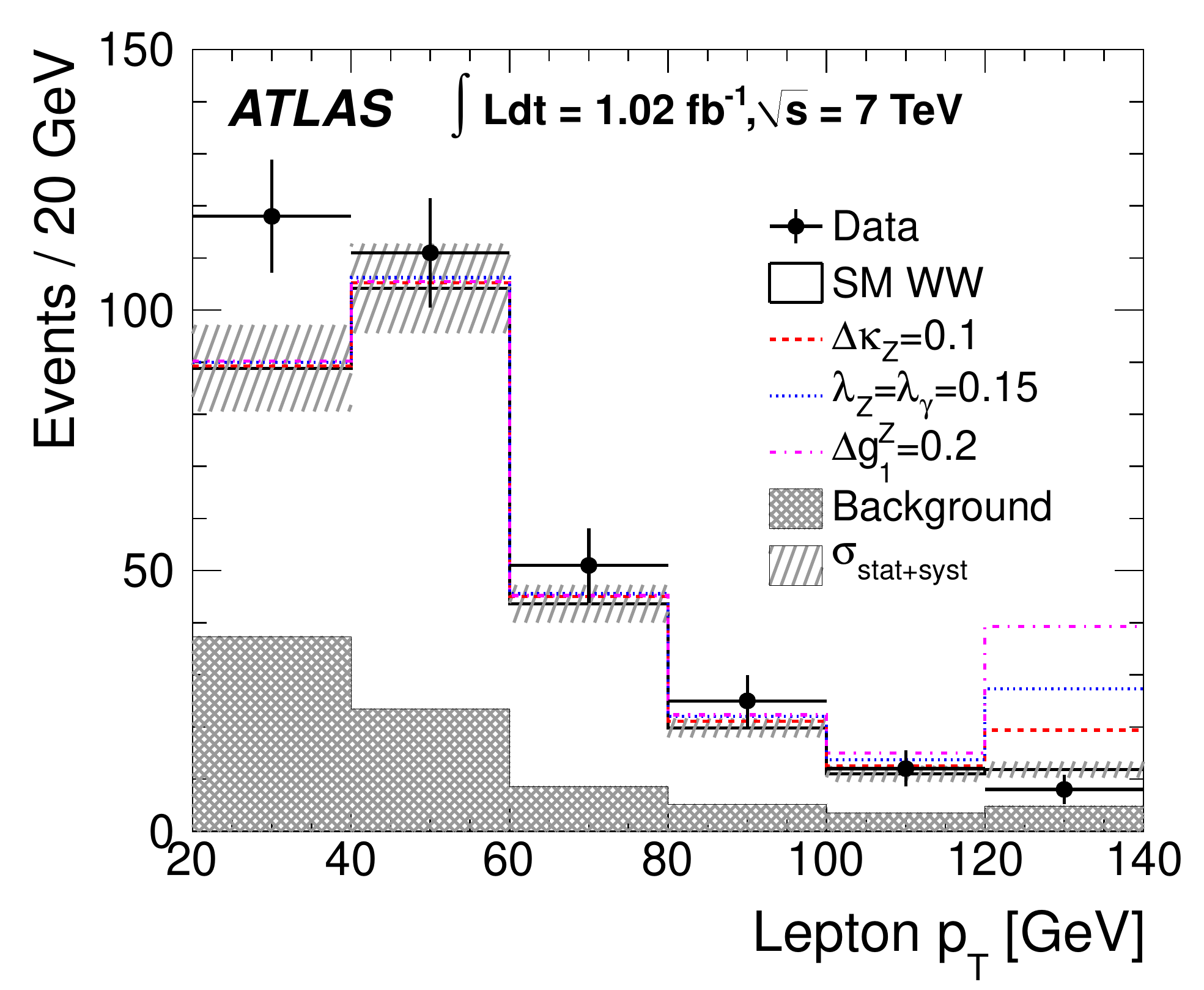}
\caption{The \pt distribution of the highest-\pt charged lepton in $WW$ final states.  
Shown are the data (dots), the background (shaded histogram), SM $WW$ plus background 
(solid histogram), and the following $WW$ anomalous couplings added to the background:  
$\Delta\kappa_Z = 0.1$ (dashed histogram), $\lambda_Z=\lambda_{\gamma}=0.15$ (dotted 
histogram), and $\Delta g_1^Z=0.2$ (dash-dotted histogram).  The last bin corresponds 
to $\pt > 120$ GeV. } 
\label{fig:leppt}
\end{center}
\end{figure}

The differential cross section as a function of the invariant mass of the $WW$ pair is 
the most direct probe of anomalous couplings, particularly at high invariant mass.  The 
mass can not be fully reconstructed but is correlated with the momentum of the individual 
leptons.  The \pt distribution of the highest-\pt charged lepton is therefore a sensitive 
probe of anomalous TGCs and is used in a binned likelihood fit to extract the values of 
the anomalous couplings preferred by the data (Fig.~\ref{fig:leppt}).  The dependence of 
the distribution on specific anomalous couplings is modeled by reweighting the {\sc mc@nlo} 
SM $WW$ MC to the predictions of the BHO generator \cite{bho} at the matrix-element level.  
Figure~\ref{fig:leppt} demonstrates the sensitivity to anomalous TGCs at high lepton \pt; 
the coupling measurement is negligibly affected by the excess in the data at low \pt.  The 
fiducial cross section is measured in the last bin of Fig.~\ref{fig:leppt}.  The result 
$\sigma_{\rm fid} (\pt \geq 120$ GeV$) = 5.6^{+5.4}_{-4.4}~{\rm (stat.)} \pm 2.9~{\rm (syst.)} 
\pm 0.2~{\rm (lumi.)}$ fb is consistent with the SM $WW$ prediction of $\sigma_{\rm fid} 
(\pt \geq 120$ GeV$) = 12.2 \pm 1.0$ fb. 

\begin{figure*}[!hbtp]
\begin{center}
\includegraphics[width=0.33\textwidth]{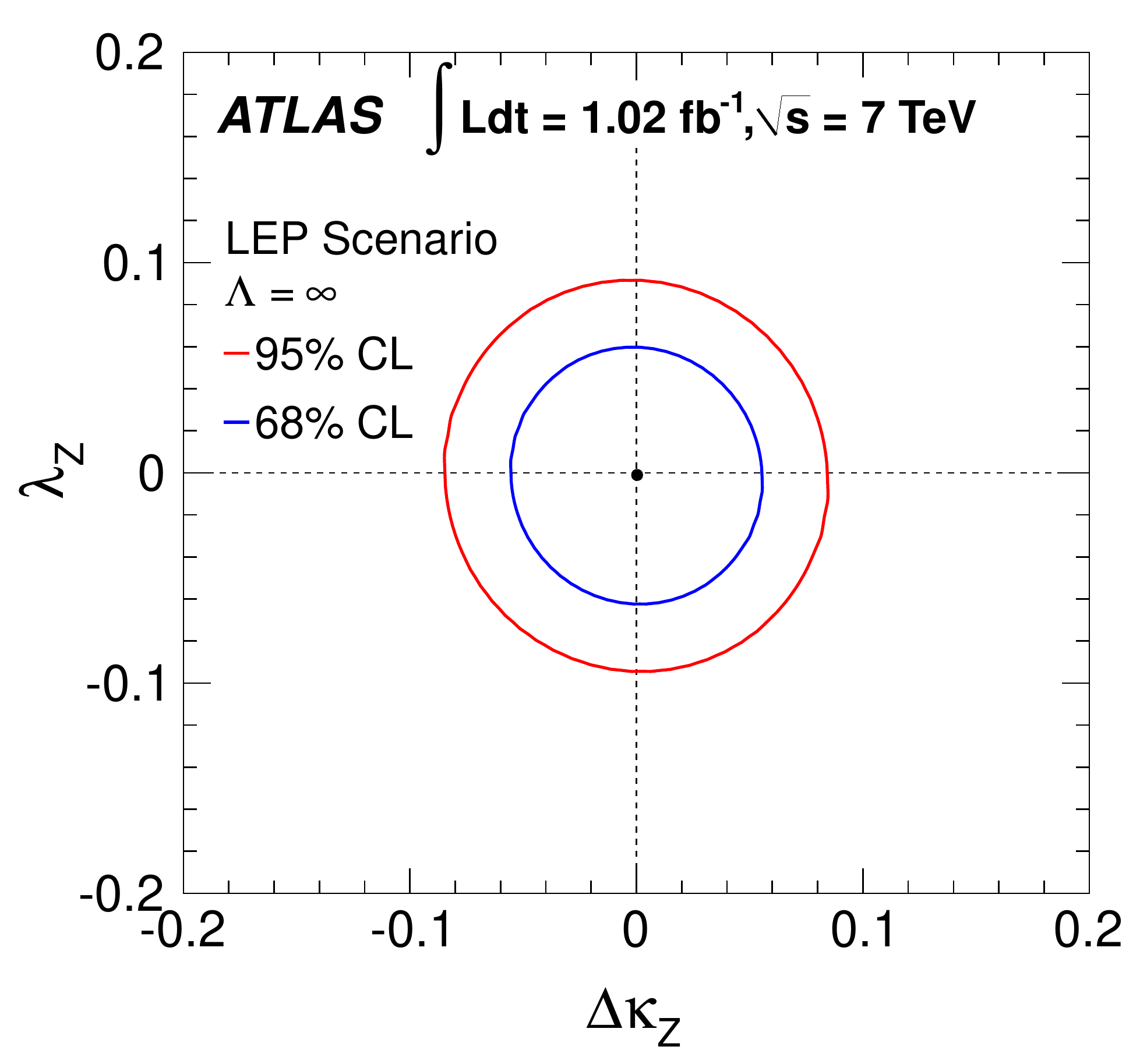}
\includegraphics[width=0.33\textwidth]{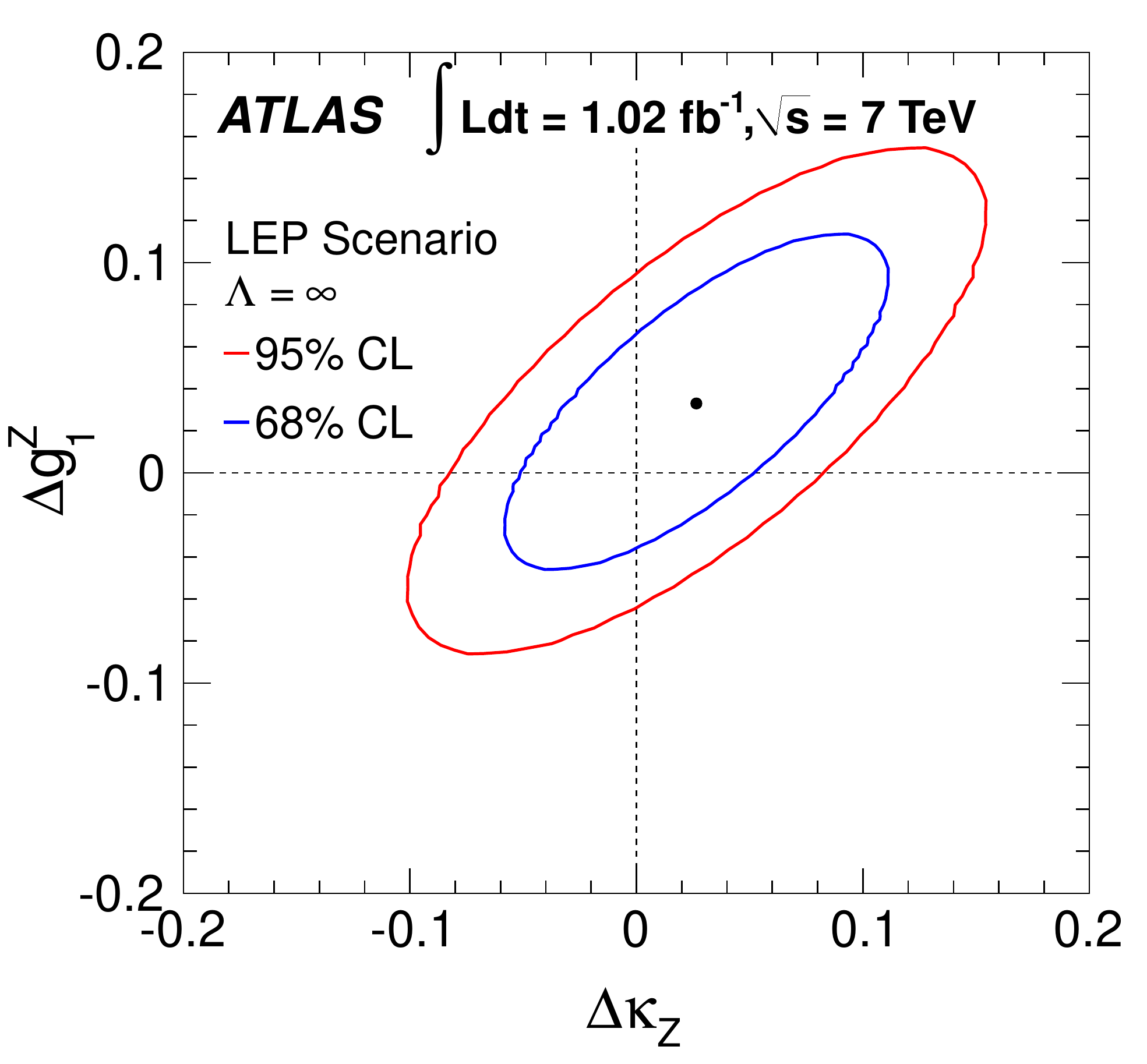}
\includegraphics[width=0.33\textwidth]{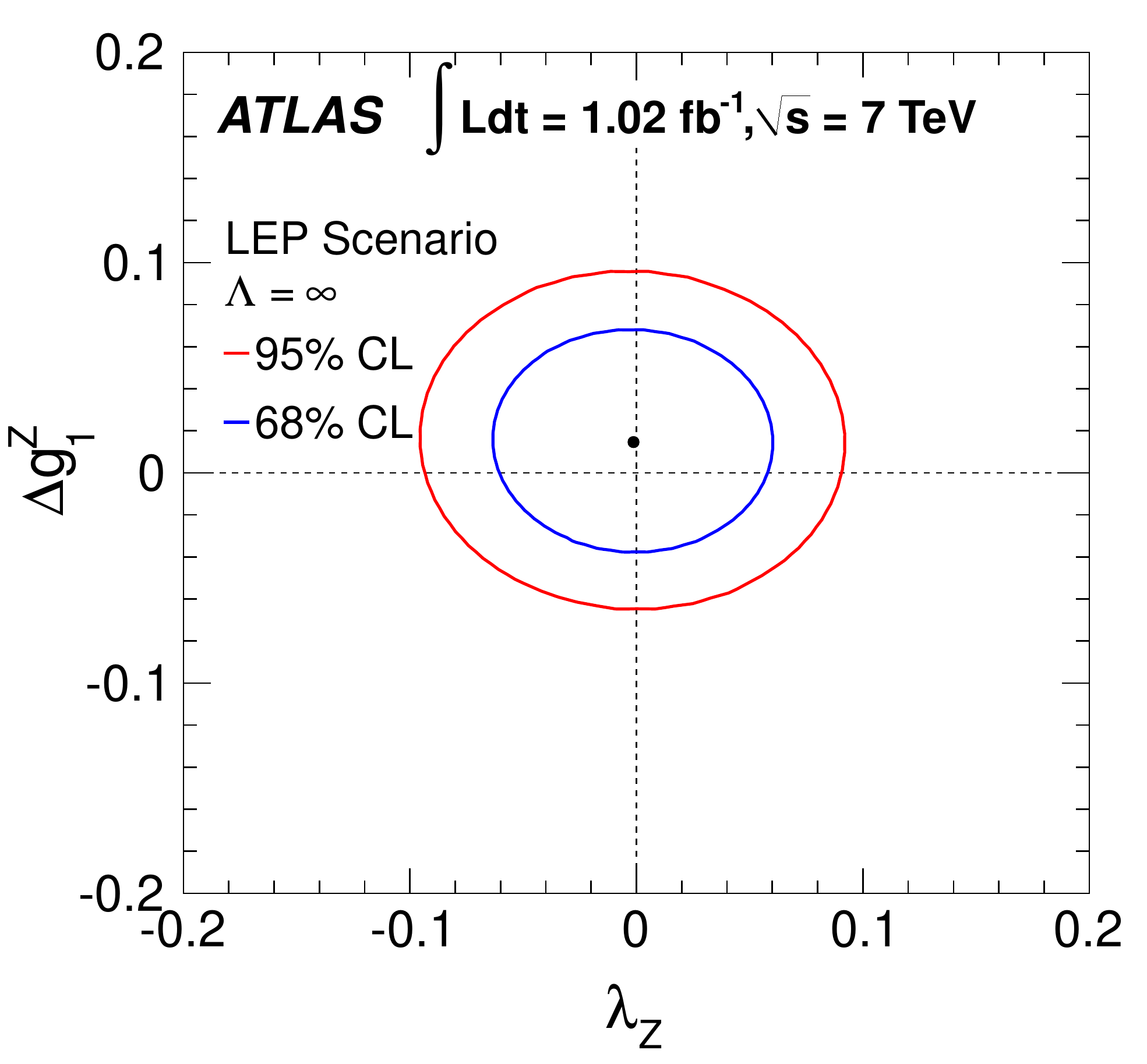}
\caption{Two-dimensional fits to the anomalous couplings in the LEP scenario: 
$\Delta \kappa_Z$ vs.~$\lambda_Z$ (left), $\Delta \kappa_Z$ vs.~$\Delta g_1^Z$ (middle), 
and $\lambda_Z$ vs.~$\Delta g_1^Z$ (right). }
\label{fig:atgc}
\end{center}
\end{figure*}

Table~\ref{tbl:atgc} and Fig.~\ref{fig:atgc} show the results of the coupling fits to 
one and two parameters respectively in the LEP scenario, with $\Lambda = 3$ TeV and the 
other parameter(s) set to the SM value(s).  One-dimensional limits on $\lambda_Z$ in 
the HISZ and Equal Coupling scenarios are the same as in the LEP scenario.  In the HISZ 
scenario, the 95\% CL limits on $\Delta\kappa_Z$ are $[-0.049,~0.072]$ and $[-0.037,~0.069]$ 
for $\Lambda = 3$ TeV and $\Lambda = \infty$, respectively.  The corresponding limits in 
the Equal Coupling scenario are $[-0.089, 0.096]$ and $[-0.065, 0.102]$, respectively.

\begin{table*}[!hbtp]
\caption{95\% CL limits on anomalous TGCs in the LEP scenario
assuming the other couplings are set to their SM values. }
\begin{center}
\begin{tabular}{cccc}
\hline 
$\Lambda$& $\Delta g_1^Z$    & $\Delta\kappa_Z$  & $\lambda_Z$ \\
\hline 
3 TeV    & $[-0.064,~0.096]$ & $[-0.100,~0.067]$ & $[-0.090,~0.086]$ \\
$\infty$ & $[-0.052,~0.082]$ & $[-0.071,~0.071]$ & $[-0.079,~0.077]$ \\
\hline
\end{tabular}
\end{center}
\label{tbl:atgc}
\end{table*}

\begin{figure*}[!tp]
\begin{center}
\includegraphics[width=0.8\textwidth]{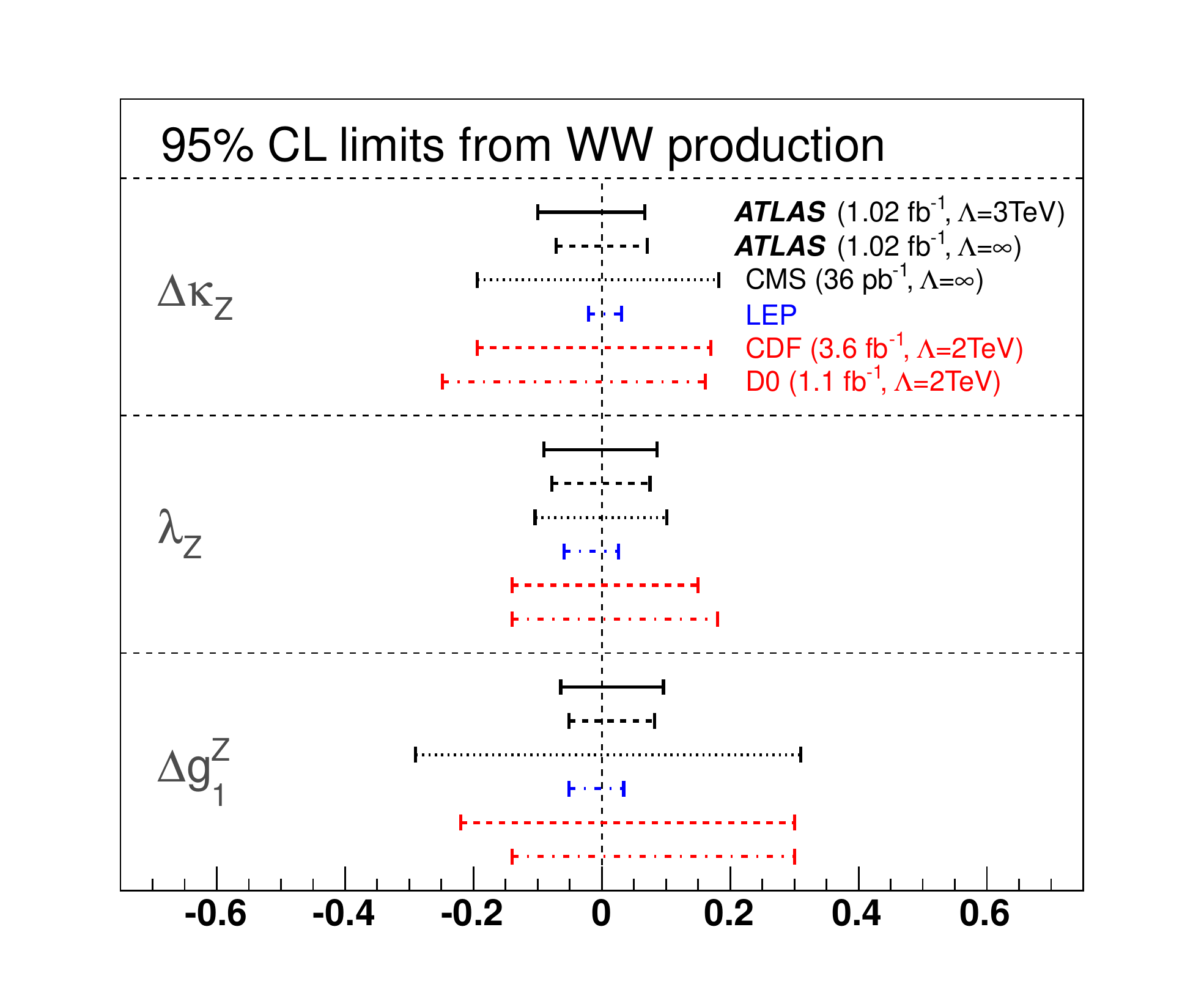}
\caption{Anomalous TGC limits from ATLAS, D0 and LEP (based on the LEP scenario) and 
CDF and CMS (based on the HISZ scenario), as obtained from $WW$ production measurements. }
\label{fig:tgc_limits}
\end{center}
\end{figure*}

The anomalous coupling limits in the LEP scenario are compared with limits obtained from 
CMS, CDF, D0 and the combined LEP results in Fig.~\ref{fig:tgc_limits}.  The sensitivity 
of this result is significantly greater than that of the Tevatron due to the higher 
center-of-mass energy and higher $WW$ production cross section.  It is also comparable to 
the combined results from LEP, which include data from four detectors and all $WW$ decay 
channels.

\section{Summary}
\label{summary}
Using 1.02 fb$^{-1}$ of $\sqrt{s} = 7$ TeV $pp$ data, the $pp\rightarrow WW$ cross section 
has been measured with the ATLAS detector in the fully leptonic decay channel.  The measured 
total cross section of $54.4 \pm 5.9$ pb is consistent with the SM prediction of 
$44.4 \pm 2.8$ pb and is the most precise measurement to date.  In addition, a first 
measurement of the $WW$ cross section in a fiducial phase space region has been presented.  
Limits on anomalous couplings have been derived in three scenarios using the \pt distribution 
of the leading charged lepton.  No significant deviation is observed with respect to the SM 
prediction.  These limits are competitive with previous results and are sensitive to a higher 
mass scale for new physical processes.

\section{Acknowledgements}
We thank CERN for the very successful operation of the LHC, as well as the support staff 
from our institutions without whom ATLAS could not be operated efficiently.

We acknowledge the support of ANPCyT, Argentina; YerPhI, Armenia; ARC,
Australia; BMWF, Austria; ANAS, Azerbaijan; SSTC, Belarus; CNPq and FAPESP,
Brazil; NSERC, NRC and CFI, Canada; CERN; CONICYT, Chile; CAS, MOST and NSFC,
China; COLCIENCIAS, Colombia; MSMT CR, MPO CR and VSC CR, Czech Republic;
DNRF, DNSRC and Lundbeck Foundation, Denmark; EPLANET and ERC, European Union;
IN2P3-CNRS, CEA-DSM/IRFU, France; GNAS, Georgia; BMBF, DFG, HGF, MPG and AvH
Foundation, Germany; GSRT, Greece; ISF, MINERVA, GIF, DIP and Benoziyo Center,
Israel; INFN, Italy; MEXT and JSPS, Japan; CNRST, Morocco; FOM and NWO,
Netherlands; RCN, Norway; MNiSW, Poland; GRICES and FCT, Portugal; MERYS
(MECTS), Romania; MES of Russia and ROSATOM, Russian Federation; JINR; MSTD,
Serbia; MSSR, Slovakia; ARRS and MVZT, Slovenia; DST/NRF, South Africa;
MICINN, Spain; SRC and Wallenberg Foundation, Sweden; SER, SNSF and Cantons of
Bern and Geneva, Switzerland; NSC, Taiwan; TAEK, Turkey; STFC, the Royal
Society and Leverhulme Trust, United Kingdom; DOE and NSF, United States of
America.

The crucial computing support from all WLCG partners is acknowledged
gratefully, in particular from CERN and the ATLAS Tier-1 facilities at
TRIUMF (Canada), NDGF (Denmark, Norway, Sweden), CC-IN2P3 (France),
KIT/GridKA (Germany), INFN-CNAF (Italy), NL-T1 (Netherlands), PIC (Spain),
ASGC (Taiwan), RAL (UK) and BNL (USA) and in the Tier-2 facilities
worldwide.

\bibliographystyle{model1-num-names}
\bibliography{<your-bib-database>}

\clearpage
\onecolumn
\begin{flushleft}
{\Large The ATLAS Collaboration}

\bigskip

G.~Aad$^{\rm 48}$,
B.~Abbott$^{\rm 112}$,
J.~Abdallah$^{\rm 11}$,
S.~Abdel~Khalek$^{\rm 116}$,
A.A.~Abdelalim$^{\rm 49}$,
A.~Abdesselam$^{\rm 119}$,
O.~Abdinov$^{\rm 10}$,
B.~Abi$^{\rm 113}$,
M.~Abolins$^{\rm 89}$,
O.S.~AbouZeid$^{\rm 159}$
% ...
\bigskip

% ...
$^{10}$ Institute of Physics, Azerbaijan Academy of Sciences, Baku, Azerbaijan\\
$^{11}$ Institut de F\'isica d'Altes Energies and Departament de F\'isica de la Universitat Aut\`onoma  de Barcelona and ICREA, Barcelona, Spain\\
% ...
$^{48}$ Fakult\"{a}t f\"{u}r Mathematik und Physik, Albert-Ludwigs-Universit\"{a}t, Freiburg i.Br., Germany\\
$^{49}$ Section de Physique, Universit\'e de Gen\`eve, Geneva, Switzerland\\
% ...
$^{89}$ Department of Physics and Astronomy, Michigan State University, East Lansing MI, United States of America\\
% ...
$^{112}$ Homer L. Dodge Department of Physics and Astronomy, University of Oklahoma, Norman OK, United States of America\\
$^{113}$ Department of Physics, Oklahoma State University, Stillwater OK, United States of America\\
% ...
$^{116}$ LAL, Univ. Paris-Sud and CNRS/IN2P3, Orsay, France\\
% ...
$^{119}$ Department of Physics, Oxford University, Oxford, United Kingdom\\
% ...
$^{159}$ Department of Physics, University of Toronto, Toronto ON, Canada\\
% ...
\end{flushleft}


\end{document}